\documentstyle[12pt]{article}
\title{Spin-Momentum Correlation (Handedness) in the Process of Four Pions
Production in the Electron-Positron  Collisions}
\author{E.~L.~Bratkovskaya$^1$,                     E.~A.~Kuraev$^1$,\\
Z.~K.~Silagadze$^2$, O.~V.~Teryaev$^1$ \vspace*{5mm}\\
$^1$ \small\em Joint Institute for Nuclear Research, 141980 Dubna,
Moscow region, Russia \\[3mm]
$^2$ \small \em Institute of Nuclear Physics, Prospect Nauki, 11,
630090, Novosibirsk, Russia}
\date{}
\begin{document}
\maketitle

\begin{abstract}
We discuss special type of polarization asymmetry (``handedness")
in the process $e^+e^-\to 4\pi$, when one of initial particles is
longitudinally polarized. The asymmetry is proportional to the degree
of polarization and to the width to mass ratio
of rho-meson. It can reach 4-5\% in some kinematical region.
Both channels $2 \pi^+ 2 \pi^-$ and $2 \pi^0 \pi^+ \pi^-$ are considered
in the framework of the effective chiral lagrangian with vector
mesons, in the energy range $\sim 1$ GeV. The corresponding total cross
sections are also calculated.
\end{abstract}

\section{Introduction}
The handedness concept, as a measure of the initial state polarization,
was first discussed in papers of O.~Nachtman and A.~V.~Efremov
\cite{1,2}. In particular, it was suggested that the polarization of
the initial parton could be established by investigating the
characteristics of the corresponding jet \cite{3,4}.  The longitudinal
polarization of quark created in $e^+e^-$ annihilation, arises due to
the interference between vector and axial amplitudes of the $Z$-boson
intermediate state.
 However, the correlation between quark polarization and jet
handedness is expected to be greatly reduced when averaging over final
phase space because of the complicated process of jet fragmentation. So
in this case one can expect the effect to be of the order of 2-3\%,
making its experimental study an elaborate task.

In this paper we consider the similar correlation in a much simpler case
of $e^+e^-$ annihilation into four pions at intermediate energies.
It allows one to study the most probable mechanism of the handedness
generation, namely, the wide resonance imaginary phase contribution.
The quantity (handedness)
\begin{eqnarray}
H=\frac{L-R}{L+R}
\label{eq1}\end{eqnarray}
where L,R are the numbers of the left handed and right handed configurations
constructed from the two pions 3-momenta and the beam direction, is not
zero when electron and positron beams are longitudinally polarized (it
turns out that the effect is also present in the case when only one of
the beams is polarized.

We suggest to choosing the same charge sign pions for the $2\pi^+2\pi^-$
channel and arrange them according to their momenta, while
any pair of pions can be taken for the $2\pi^0\pi^+\pi^-$ channel.

Note that for $e^+e^-\to 2\pi$ and $e^+e^-\to 3\pi$ reactions the
handedness effect is absent $H=0$. This is due to the fact that in these
cases we actually have only one amplitude and can therefore measure
only the symmetrical part of the initial state spin-density matrix.

In the case of four pions production we have two types of amplitudes
which depend differently on the initial polarization and the interference
between them just gives the helicity-dependent term. Note that this
interference term is due to the nonzero width of the $\rho$-meson in some
intermediate state and the large value of this width suggests that the
considerable effect could be expected.

\section{General considerations}

The main contribution to the $e^+e^-\to 4\pi$ cross section goes from
the annihilation channel. Using the vector dominance model, the
corresponding matrix element can be presented as
\begin{eqnarray}
M^{e^+e^-\to 4\pi} = {4\pi\alpha m_\rho^2 \over s (s-m_\rho^2
+ i m_\rho \Gamma_\rho)} \ \overline v (\lambda_+, p_+)\  \gamma_\mu\
u (\lambda_-, p_-) \ J_\mu (\rho^0 \to 4 \pi),
\label{eq2}\end{eqnarray}
where $s=(p_+ + p_-)^2, \lambda_+=-\lambda_-=\pm 1$ are the initial
state positron and electron chiralities and $g_{\rho\pi\pi}
J_\mu(\rho\to 4\pi)$  is the conserved current:
\begin{eqnarray}
q_{\mu}\ J^{\mu} = 0, \ \ q=p_+ + p_-,
\label{eq3}\end{eqnarray}
which describes the $\rho\to 4\pi$ transition. Its concrete form is of
course model dependent. In the energy region $\sqrt{s}\sim 1$~GeV,
which we are going to consider, the effective chiral lagrangian with
vector mesons can give a reasonable approximation \cite{6}. We will use
the version \cite{7,8} of such an effective chiral lagrangian, which
correctly incorporates a phenomenologically successful vector meson
dominance picture \cite{9} and current algebra low energy theorems.
For convenience, let us reproduce here its relevant part:
\begin{eqnarray}
{\cal L} &=& {1 \over 2} \ {\rm Sp} \ (D_\mu \Phi) (D^\mu \Phi) +
{1 \over 2 f_\pi^2} \ ({1\over 3}-\alpha_k) {\rm Sp} \left[
\Phi \ (D_\mu \Phi) \ \Phi \ (D^\mu \Phi) - \Phi^2 \ (D_\mu \Phi) (D^\mu
\Phi)\right] \nonumber\\
&-& {\varepsilon^{\mu\nu\lambda\sigma}\over \pi^2} \left\{{3\over 8\sqrt{2}}
{g_{\rho\pi\pi}^2\over f_\pi} \ {\rm Sp} \left[ (\partial_\mu V_\nu)
(\partial_\lambda V_\sigma) \Phi\right] + i {g_{\rho\pi\pi}\over 4 f_\pi^3}
(1-3 \alpha_k) \ {\rm Sp} \left[V_\mu(\partial_\nu \Phi)
(\partial_\lambda \Phi) (\partial_\sigma \Phi)\right] \right\} \nonumber\\
&-& {e m_\rho^2\over g_{\rho\pi\pi}} A_\mu \rho^\mu - {1\over 4} {\rm Sp}
F_{\mu\nu}^{(V)} F^{(V) \mu\nu},
\label{eq4}\end{eqnarray}
where $\displaystyle\alpha_k={g_{\rho\pi\pi}^2 f_\pi^2\over
m_\rho^2}\simeq 0.55, \ D_\mu\Phi = \partial_\mu\Phi -
i {g_{\rho\pi\pi}\over \sqrt{2}} [V_\mu,\Phi], \ F_{\mu\nu}^{(V)} =
\partial_\mu V_\nu - \partial_\nu V_\mu - i {g_{\rho\pi\pi}\over \sqrt{2}}
[V_\mu,V_\nu],$ $f_\pi\simeq 93$~MeV and $\Phi, V_\mu$ are the
conventional SU(3) matrices for pseudoscalar and vector meson fields.

{}From (\ref{eq2}) we get
\begin{eqnarray}
\mid M\mid^2 = {(4\pi\alpha m_\rho^2)^2\over s^2 \left[(s - m_\rho^2)^2
+ m_\rho^2 \Gamma_\rho^2\right]} \ L_{\mu\nu}\ J^\mu \ J^{\nu\dag},
\label{eq5}\end{eqnarray}
where $L_{\mu\nu}$ lepton tensor has only transversal to beam components
in the center of mass system:
\begin{eqnarray}
&&L_{\mu\nu} = {1\over 4} \ {\rm Sp}\  \hat p_- (1 +\lambda_- \gamma_5)
\gamma_\mu \hat p_+ (1 - \lambda_+ \gamma_5) \gamma_\nu
 = {s\over 2}\ \left[ (1 - \lambda_+ \lambda_-) \delta_{\mu\nu}^\perp
+ i (\lambda_- -\lambda_+) \varepsilon_{\mu\nu}^\perp \right], \nonumber \\
&&  \delta_{\mu\nu} = diag(0,1,1,0), \ \varepsilon_{\mu\nu}^\perp =
\varepsilon_{03\mu\nu}, \ \varepsilon_{0123}=1.
\label{eq6}\end{eqnarray}
Due to the presence of the nonzero imaginary part of the $\rho\to 4\pi$
amplitude, $J_\mu J_\nu^{\dag}$ tensor has an antisymmetrical part:
\begin{eqnarray}
J_\mu J_\nu^{\dag} = (a + i b)_\mu \ (a - i b)_\nu =
a_\mu a_\nu + b_\mu b_\nu + a_\mu a_\nu + i (b_\mu a_\nu - a_\mu b_\nu).
\label{eq7}\end{eqnarray}
As a result we obtain for the cross section:
\begin{eqnarray}
&& d\sigma^{e^+e^-\to 4\pi} = {(\alpha m_\rho^2)^2 {\cal F} \over
2^6 \pi^6 s^2 \left[(s - m_\rho^2)^2 + m_\rho^2 \Gamma_\rho^2)\right]}
\ \prod_{i=1}^4 \ {d \vec q_i \over 2 E_i}\
\delta^4 (q-\sum\limits_{i=1}^4 q_i),\nonumber\\[2mm]
&& {\cal F} = (1 - \lambda_+ \lambda_-) \ \left( a_x^2 + a_y^2 + b_x^2 +
b_y^2 + 2 (\lambda_- - \lambda_+) (\vec a \times \vec b)_z \right),
\label{eq8}\end{eqnarray}
where it is assumed that z-axis coincides with the $\vec p_-$-direction.

Performing the phase space integration, one may obtain $\sigma_{L,R}$
in the form
\begin{eqnarray}
\sigma_{L,R} = {1\over 2} (1 - \lambda_+\lambda_-)\sigma_0 \pm
{1\over 2} (\lambda_- - \lambda_+) {\Gamma_\rho\over m_\rho} \sigma_1.
\label{eq9}\end{eqnarray}
In fact $\sigma_0$ is an unpolarized cross section and $\sigma_1$ is
related to the spin-momentum correlation (handedness):
\begin{eqnarray}
H = {\sigma_L - \sigma_R\over \sigma_L + \sigma_R} =
{\Gamma_\rho \over m_\rho} {\lambda_- - \lambda_+
\over  1 - \lambda_+\lambda_-} {\sigma_1\over \sigma_0}
\label{eq10}\end{eqnarray}

\section{Four charged pions production}

The types of Feynman diagrams for transition $\rho \to 2 \pi^+ 2 \pi^-$:
\begin{eqnarray}
e^+(p_+) + e^-(p_-) \to \pi^+(q_1) + \pi^+(q_2) + \pi^-(q_3)+\pi^-(q_4)
\label{eq11}
\end{eqnarray}
are shown in Fig.~1. The corresponding current has a form:
\begin{eqnarray}
&& J_\mu^{\rho^0\to 2\pi^+ 2\pi^-} = ({1\over 3}-\alpha_k) {1\over f_\pi^2}
\left[ 6 (q_1 + q_2 - q_3 - q_4)_\mu + (6 q_3.q_4 + 2 m^2)
\left({(q - 2 q_1)_\mu\over (q - q_1)^2 - m^2} \right.\right. \nonumber\\
&&+ \left.\left. {(q - 2 q_2)_\mu\over
(q - q_2)^2 - m^2}\right) - ( 6 q_1.q_2 + 2 m^2)\left({(q - 2q_3)_\mu
\over (q - q_3)^2 - m^2} + {(q - 2 q_4)_\mu\over (q - q_4)^2 - m^2}
\right)\right]  \nonumber\\
&& + 2 (1 + P_{12}) (1 + P_{34}) {g_{\rho\pi\pi}^2((q_2 + q_4)^2 - m_{\rho}^2
- i m_\rho \Gamma_\rho) \over ((q_2 + q_4)^2 - m_\rho^2)^2
+ m_\rho^2\Gamma_\rho^2} \nonumber\\
&&\times \left[ (q_4 - q_2)_\mu+ {q_1.(q_2 - q_4)\over
(q - q_3)^2 - m^2} (q - 2 q_3)_\mu + {q_3.(q_2 - q_4)\over (q - q_1)^2
- m^2} (q - 2 q_1)_\mu\right],
\label{eq12}
\end{eqnarray}
where $m^2=m_\pi^2=q_i^2$. $P_{12}$ and $P_{34}$ operators stand for
the interchange of the corresponding identical mesons momenta.

Consider now the equally charged pions ($\pi^+$ for example), arranged
according to the magnitude of their momenta (say, more energetic particle
defines
an x-axis direction), and let them together with the beam axis (for
definiteness $\vec p_-$) form a left or right configurations.
The numbers of the left and right repers will not in general coincide,
if the initial state is characterized by some nonzero average longitudinal
polarization. The corresponding assymmetry (handedness) is
given by (\ref{eq10}). Using a standard covariant phase-space calculations
\cite{10}, (\ref{eq8}) and (\ref{eq9}) can be cast in the following form
\begin{eqnarray}
\sigma_{0,1} = {(\alpha m_\rho^2)^2 \over 2^7 \pi^6 s^2}
{R_{0,1}\over \left[(s - m_\rho^2)^2 + m_\rho^2 \Gamma_\rho^2)\right]},
\label{eq13}\end{eqnarray}
where
\begin{eqnarray}
R_0={\pi^2\over 24 s} \int\limits_{s_1^-}^{s_1^+} d s_1
\int\limits_{s_2^-}^{s_2^+} d s_2  \int\limits_{u_1^-}^{u_1^+}
{d u_1 \over \sqrt{\lambda(s,s_2,s_2^\prime)}}  \int\limits_{u_2^-}^{u_2^+}
d u_2 \int\limits_{-1}^1 {d\zeta \over \sqrt{1-\zeta^2}} |\vec J|^2
\label{eq14}\end{eqnarray}
(the expressions for the integration limits, as well as some details of
calculations are given in the appendix).
Assuming that $\vec J$ from (\ref{eq12}) is presented as
\begin{eqnarray}
\vec J = D_1 \vec q_1 + D_2 \vec q_2 + D_3 \vec q_3,
\label{eq15}\end{eqnarray}
$R_1$ is given by a similar expression
\begin{eqnarray}
R_1={\pi^2\over 8 s} \int\limits_{s_1^-}^{s_1^+} d s_1
\int\limits_{s_2^-}^{s_2^+} d s_2  \int\limits_{u_1^-}^{u_1^+}
d u_1 {\theta(u_1-s_1) \over \sqrt{\lambda(s,s_2,s_2^\prime)}}
\int\limits_{u_2^-}^{u_2^+} d u_2 \int\limits_{-1}^1 {d\zeta \over
\sqrt{1-\zeta^2}} \sqrt{{\Delta_3 (q,q_1,q_2) \over s}} \left({m_\rho\over
\Gamma_\rho } \ f_1 \right),
\label{eq16}\end{eqnarray}
where
\begin{eqnarray}
f_1={i\over 2} (D_1 D_2^\star - D_2 D_1^\star)
\label{eq17}\end{eqnarray}
and
\begin{eqnarray}
\Delta_3(q,q_1,q_2) &=& \left|\begin{array}{ccc}
q.q     &  q.q_1    & q.q_2   \\
q_1.q   &  q_1.q_1  & q_1.q_2 \\
q_2.q   &  q_2.q_1  & q_2.q_2
\end{array} \right| \nonumber\\[3mm]
&=& \left|\begin{array}{ccc}  \displaystyle
 s     &  \displaystyle  {1\over 2} (s+m^2-s_1)
       & \displaystyle  {1\over 2} (s+m^2-u_1)  \\[2mm]
 \displaystyle {1\over 2} (s+m^2-s_1) & \displaystyle  m^2
       & \displaystyle  {1\over 2} (s+s_2-s_1-u_1)  \\[2mm]
 \displaystyle {1\over 2} (s+m^2-u_1)   &
\displaystyle {1\over 2} (s+s_2-s_1-u_1) &  \displaystyle m^2
\end{array}\right|
\label{eq18}\end{eqnarray}
$\theta(u_1 - s_1)$ in (\ref{eq16}) is equivalent to $\theta(E_1-E_2)$
and expresses an arrangement of identical pions according to their energy.

The results of the numerical calculations are presented in Fig.~2.
The unpolarized total cross section $\sigma_0$ is also shown in
Fig.~3 together with the experimental data \cite{11}.

\section{ $2 \pi^0 \pi^+ \pi^-$-- channel}

For the process
\begin{eqnarray}
e^+(p_+) + e^-(p_-) \to \pi^+(q_+) + \pi^-(q_-) + \pi^0(q_1) + \pi^0(q_2)
\label{eq19}\end{eqnarray}
some additional Feynman diagrams with the vertices from the anomalous
part of chiral Lagrangian are essential. The types of relevant diagrams are
drawn
in Fig.~1. The corresponding current $J_\mu$ can be presented as a sum
of three terms, each representing a gauge invariant subset
of diagrams:
\begin{eqnarray}
J_\mu^{\rho\to 2\pi_0 \pi_+ \pi_-} = J_\mu^{(1)} + J_\mu^{(2)} + J_\mu^{(3)}.
\label{eq20}\end{eqnarray}
Diagrams of type a,b of Fig.~1 give:
\begin{eqnarray}
J_\mu^{(1)} = ({1\over 3} -\alpha_k) {1\over f_\pi^2}
(6 q_1.q_2 + 2 m_{\pi^0}^2)\ \left[{(q - 2 q_-)_\mu\over (q - q_-)^2 -
m_{\pi^\pm}^2}\ - {(q - 2 q_+)_\mu \over (q - q_+)^2 - m_{\pi^\pm}^2}\right].
\label{eq21} \end{eqnarray}
The second piece arises from diagrams of type c,d,e of Fig.~1 and has
the form
\begin{eqnarray}
J_\mu^{(2)} &=& - g_{\rho\pi\pi}^2\, (1+P_{12})\, \left\{ -{1\over r_+r_-}
\left[ 2 (q_+ - q_1)_\mu q.(q_- - q_2) - 2 (q_- - q_2)_\mu q.(q_+ - q_1)
\right.\right.\nonumber\\
&+& \left.(q_2 + q_- - q_1 - q_+)_\mu (q_+ - q_1).(q_- - q_2)\right]
\nonumber\\
&+& {1\over r_+}\, \left[ (q_+ - q_1)_\mu - 2 q_2.(q_+ - q_1)
{(q - 2 q_-)_\mu \over (q - q_-)^2 - m_\pi^2}\right] \nonumber\\
&-&\left.  {1\over r_-}\, \left[ (q_- - q_2)_\mu - 2 q_1.(q_- - q_2)
{(q - 2 q_+)_\mu \over (q - q_+)^2 - m_\pi^2 } \right] \right\},
\label{eq22}\\[3mm]
r_+ &=& (q_+ +q_1)^2 - m_\rho^2 + i m_\rho \Gamma_\rho; \ \ \
r_- = (q_- + q_2)^2 - m_\rho^2 + i m_\rho \Gamma_\rho.
\nonumber
\end{eqnarray}
Finally, the third part of current is determined by two diagrams of type f
of Fig.~1 with the $\omega$-meson intermediate state:
\begin{eqnarray}
J_\mu^{(3)} = {3 g_{\rho\pi\pi} \over 8\pi^2 f_\pi}\, (1+P_{12})\,
P_\mu \, {F_1\over r_1},
\label{eq23} \end{eqnarray}
where
\begin{eqnarray}
P_\mu = q_1.q_2 (q_{+\mu} q.q_- - q_{-\mu} q.q_+) + q_-.q_2 (q_{1\mu}
q.q_+ - q_{+\mu} q.q_1) + q_+.q_2 (q_{-\mu} q.q_1 - q_{1\mu} q.q_-),
\label{eq24}\end{eqnarray}
and
\begin{eqnarray}
&&r_1 = (q-q_2)^2 - m_\omega^2 + i m_\omega \Gamma_\omega, \nonumber\\[2mm]
&&F_1 = {3g_{\rho\pi\pi} \over 4 \pi^2 f_\pi^3}
\left[1-3\alpha_k -\alpha_k\left({m_\rho^2\over r_{+-}} +
{m_\rho^2\over r_{+1}} + {m_\rho^2\over r_{-1}} \right)\right],\nonumber\\[2mm]
&& r_{+-} = (q_+ + q_-)^2 - m_\rho^2 + i m_\rho \Gamma_\rho, \nonumber\\
&& r_{+1} = (q_+ + q_1)^2 - m_\rho^2 + i m_\rho \Gamma_\rho, \nonumber\\
&& r_{-1} = (q_1 + q_-)^2 - m_\rho^2 + i m_\rho \Gamma_\rho. \label{eq25}
\end{eqnarray}
The handedness value in the case when two $\pi^0$'s are taken to define
a reper is less a 1
At last, in Fig.~4 we draw the calculated total unpolarized cross section
compared to the experimental data from \cite{11}.

The known experimental data for $\sqrt{s} < 1$~GeV \cite{11} are in
resonable agreement with our calculation  of total cross section for
$2\pi^0\pi^+\pi^-$ channel. In calculations, we have taken into account the
dependence of the $\rho$-meson width on energy.
The situation is worse for $\sqrt{s} > 1$~GeV. For $\sqrt{s} = 1.3$~GeV
the experimental cross section exceeds about one order of magnitude  the ones
obtained above (\ref{eq8}) (See Fig.~4). Presumably, the difference arises
mainly from the influence of the $\rho$-meson radial exitation ---
$\rho^\prime$ (1450) resonance. Let us now introduce an additional
factor $R(s)$ in the cross section $d\sigma(s)\to d\sigma(s) R(s)$,
\begin{eqnarray}
R(s) = \left|{m_\rho^2\over s-m_\rho^2+i m_\rho\Gamma_\rho}\right|^{-2}
\cdot \left|{m_\rho^2\over s-m_\rho^2+i m_\rho\Gamma_\rho} +
{m_{\rho^\prime}^2 e^{i\varphi} \over s-m_{\rho^\prime}^2+i m_{\rho^\prime}
\Gamma_{\rho^\prime}} \right|^2,
\label{eq26}\end{eqnarray}
which takes into account the $\rho^\prime$-meson contribution. From
Figs.~3,5 we see that it works in the useful direction. Note in conclusion,
that the value of handedness (\ref{eq10}) will not be changed after the
replacement $d\sigma\to R d\sigma$.

As for $a_1$-meson contribution, for
low energy region $\sqrt{s} < 1$ GeV it is effectively taken
into account via the effective coupling constant in the lagrangian.
This obviously becomes incorrect when $\sqrt{s} > 1.3$ GeV, where
$3\pi$-invariant mass can reach such values that
the Breit-Wigner character of $a_1$-meson intermediate state propagator
is essential.
It's the resons why the above given formulas can not be applied in
the  $\sqrt{s} > 1.3$ GeV region.

\section*{Acknowledgments}
We are grateful to A.~V.~Efremov and S.~I.~Eidelman for critical
reading and fruitful discussions.

\appendix
\section{Appendix. Covariant phase-space calculations}
\def\theequation{\thesection.\arabic{equation}}
\setcounter{equation}{0}

Let us consider
\begin{eqnarray}
R_4 = \int |{\cal M}|^2 \delta(q - \sum\limits_{j=1}^4 q_j)
\prod\limits_{i=1}^4 {d\vec q\over 2 E_i}.
\label{eqA1}\end{eqnarray}
If we introduce Kumar's invariant variables
\begin{eqnarray}
s_1=(q-q_1)^2,  s_2=(q-q_1-q_2)^2,  u_1=(q-q_2)^2, \ u_2=(q-q_3)^2,
t_2=(q-q_2-q_3)^2,
\label{eqA2}\end{eqnarray}
(\ref{eqA1}) can be recasted in the form \cite{10} (assuming that $|M|^2$
is rotational invariant):
\begin{eqnarray}
R_4={\pi^2\over 8 M^2} \int\limits_{s_1^-}^{s_1^+} d s_1
\int\limits_{s_2^-}^{s_2^+} d s_2  \int\limits_{u_1^-}^{u_1^+}
d u_1  \int\limits_{u_2^-}^{u_2^+} d u_2 \int\limits_{-1}^1
{d\zeta \over \sqrt{1-\zeta^2}} {|{\cal M}|^2\over
\sqrt{\lambda(s,s_2,s_2^\prime)}}
\label{eqA3}\end{eqnarray}
where $s_2^\prime = s_2+s+m_1^2+m_2^2-u_1-s_1$ and ${\rm arccos}\zeta$ is an
angle between $(\vec q_2, \vec q_1 + \vec q_2)$ and $(\vec q_3, \vec q_1 +
\vec q_2)$ planes. $\lambda(x,y,z)=(x+y+z)^2-4xy$ is a conventional triangle
function. $t_2$ and $\zeta$ are related by
\begin{eqnarray}
t_2&=&m_3^2+u_1-{(s+u_1-m_2)^2(s+m_3^2-u_2)\over 2s} -
{\{ \lambda(s,m_2^2,u_1) \lambda(s,m_3^2,u_2)\}^{1/2}\over 2s}\nonumber\\
&\times& (\xi\eta-\zeta\sqrt{(1-\xi^2)(1-\eta^2)}),
\label{eqA4}\end{eqnarray}
${\rm arccos}\xi$ and ${\rm arccos}\eta$ being angles, respectively, between
$\vec q_2$ and $\vec q_1 + \vec q_2$ vectors, and $\vec q_3$ and
$\vec q_1 + \vec q_2$ vectors. They can be expressed by invariant
variables (\ref{eqA2}) as follows \cite{10}:
\begin{eqnarray}
&&\xi = {\lambda(s,s_2,s_2^\prime)+\lambda(s,m_2^2,u_1)-\lambda(s,m_1^2,s_1)
\over 2 \{\lambda(s,s_2,s_2^\prime) \lambda(s,m_2^2,u_1)\}^{1/2}} \nonumber\\
&& \eta = {\lambda(s,m_4^2,s_3^\prime)- \lambda(s,s_2,s_2^\prime)-
\lambda(s,m_3^2,u_2)\over \{
\lambda(s,s_2,s_2^\prime) \lambda(s,m_3^2,u_2)\}^{1/2}},
\label{eqA5}\end{eqnarray}
where $s_3^\prime = 2s+\sum\limits_{i=1}^4 m_i^2 - s_1 - u_1 - u_2$.
The limits of integration for $s$-type variables are
\begin{eqnarray}
s_1^- = (m_2+m_3+m_4)^2, \ s_1^+ = (\sqrt{s}-m_1)^2, \ s_2^- =(m_3+m_4)^2,
\ s_2^+ = (\sqrt{s_1} - m_2)^2.
\label{eqA6}\end{eqnarray}
While the limits for $u$-type variables are defined from $|\xi|<1,
|\eta|<1$ and look like
\begin{eqnarray}
&&\!\!\!\!\!\!\!\!\!\!\!  u_1^\pm = s+m_2^2-{(s_1+m_2^2-s_2)
(s+s_1-m_1^2)\over 2s_1} \pm
{\{\lambda(s_1,m_2^2,s_2)\lambda(s,s_1,m_1^2)\}^{1/2}\over 2s_1}\nonumber\\
&&\!\!\!\!\!\!\!\!\!\!\!  u_2^\pm = s+m_3^2-{(s_2+m_3^2-m_4^2)
(s+s_2-s_2^\prime)\over 2s_2} \pm
{\{\lambda(s_2,m_3^2,m_4^2)\lambda(s,s_2,s_2^\prime)\}^{1/2}\over 2s_2}
\label{eqA7}\end{eqnarray}
If we use (\ref{eqA3}) and note that for $\sigma_0$, $|{\cal M}|^2=|J_x|^2+
|J_y|^2$ can be replaced by ${2\over 3} |\vec J|^2$ and $u_1>s_1$
condition, which is assumed when calculating $\sigma_{L,R}$, can be
omitted and replaced by a factor ${1\over 2}$, we recover (\ref{eq14})
formula.

Dealing with $\sigma_1$  more care is needed when integrating over
$\vec q_1$ and $\vec q_2$ angular variables. It is assumed in (\ref{eqA3})
that $|{\cal M}|^2$ doesn't depend from three of them and so these
integrations give $8\pi^2$. This is no longer true in the case of $\sigma_1$,
because now $|{\cal M}|^2=2(\vec a\times \vec b)_z$. After integrating over
$d\vec q_3$, this can be replaced by $|{\cal M}|^2=f_1(\vec q_1\times
\vec q_2)_z$. Let us choose the following system for $d\vec q_2$ integration:
$z$-axis is along $\vec q_1$ and $\vec p_-$ vector lies in the $x,z$-plane,
than $(\vec q_1\times \vec q_2) \cdot \hat{\vec p}_- = -|\vec q_1| |\vec q_2|
\sin\theta_1 \sin\theta_2\sin\varphi_2$. Left or right reper means
$(\vec q_1\times \vec q_2)\cdot \hat{\vec p}_- > 0 $ or
$(\vec q_1\times \vec q_2)\cdot \hat{\vec p}_- < 0 $ and so
$\pi\le \varphi_2 \le 2\pi$ for left configuration and
$0\le \varphi_2 \le \pi$ for right one. Therefore the integration over
$d\varphi_2$ gives $\pm 2|\vec q_1| |\vec q_2| \sin\theta_1 \sin\theta_2$.
The integration over $d\Omega_1=\sin\theta_1 d\theta_1 d\varphi_1$ now
gives a factor $\pi^2$.  So the net effect of these integrations is the
change $|{\cal M}|^2\to \pm 2\pi^2 |\vec p_1| |\vec p_2| \sin\theta_2$.
It can be checked \cite{10} that $\displaystyle
\sin\theta_2=\sin\theta_{12}~=~{1\over |\vec p_1| |\vec p_2|}
\sqrt{\Delta_3(q,q_1,q_2)\over s}$, and so we recover the result for
$\sigma_1$ cited in the text.

\newpage
\begin{center}{\bf Figure captions}\end{center}

\noindent
Fig.~1. The Feynman diagrams for the process $e^+e^-\to 4 \pi$.\\[5mm]

\noindent
Fig.~2. The handedness value in the case
$e^+ + e^- \to \pi^+ + \pi^+ + \pi^- +\pi^-$.\\[5mm]

\noindent
Fig.~3. The unpolarized total cross section $\sigma_0$ for the process
$e^+ + e^- \to \pi^+ + \pi^+ + \pi^- +\pi^-$.
The experimental data are taken from Ref.\cite{11}.
Dashed line --- $\rho^\prime$ meson is added according to (\ref{eq26})
with $\varphi=180^o$.\\[5mm]

\noindent
Fig.~4. The unpolarized total cross section $\sigma_0$ for the process
$e^+ + e^- \to \pi^0 + \pi^0 + \pi^+ +\pi^-$.
The experimental data are taken from Ref.\cite{11}.
Dashed line --- $\rho^\prime$ meson is added according to (\ref{eq26})
with $\varphi=180^o$.

\end{document}